\documentclass[10pt]{article}
\usepackage{amssymb}
\usepackage{amsmath}
\usepackage{wasysym}
\usepackage{graphicx}
\begin{document}
	
	\title{Born's rule deviations from temporal non-local effects}
	
	\author{C. Dedes\thanks{{c\_dedes@yahoo.com}}\\
\\ Veritas Education, \\
The Leadenhall Building, \\
EC3V 4AB, \\
70 Gracechurch St,\\
United Kingdom \\ }

	\maketitle

	\begin{abstract}

   We investigate deviations from Born's rule in quantum systems where the quantum-equilibrium hypothesis, $\rho \neq |\Psi|^2$, fails. Using the quantum-hydrodynamic framework, we show that transit-interference phenomena and intrinsic memory effects induce finite-time non-local correlations, resulting in violations of Born's rule. These effects appear as non-exponential decay in the survival probability of unstable states at intermediate timescales. Our findings challenge conventional interpretations of quantum probability and suggest novel dynamics that could guide future experimental investigations of temporal non-locality.

	\end{abstract}

Born's rule, a cornerstone of quantum mechanics, identifies the probability density $\rho$ with the squared modulus of the wavefunction, $|\Psi|^2$. Traditionally, attempts to modify quantum predictions focus on altering the Schr\"{o}dinger equation, often by introducing nonlinear terms. In contrast, this work pursues a different strategy: we retain the Schr\"{o}dinger equation in its standard form and instead investigate potential deviations from Born's rule.

The Bohmian formulation of quantum theory offers a natural framework for this exploration, as it distinguishes between $\rho$ and $|\Psi|^2 $ at the logical level, even though they coincide in the quantum-equilibrium state. To capture deviations, we derive a deterministic second-order differential equation for the probability density from the standard continuity equation, thereby incorporating finite-time nonlocal effects. Within the quantum-hydrodynamic framework, we show that transit-interference phenomena \cite{Moshinsky1952,Kleber1994} and intrinsic memory effects drive these nonlocal correlations, giving rise to violations of Born's rule. Such effects appear as non-exponential decay in the survival probability of unstable states at intermediate timescales. By questioning the universality of Born's rule, our results open new theoretical avenues for understanding temporal nonlocality and suggest experimental pathways for probing non-equilibrium quantum dynamics.

To establish the theoretical foundation for our investigation of deviations from Born's rule, we begin by introducing the time-dependent Schr\"{o}dinger equation for a single particle in the non-relativistic regime, expressed as
	
	\begin{equation}
	i\hbar\frac{\partial \Psi (x,t)}{\partial t} = 
	\left[-\frac{\hbar ^{2}}{2m}\nabla ^{2}+V(x,t)\right]\Psi (x,t), 
	\end{equation}
	
	\noindent
	where $V(x,t)$ denotes the time-dependent external potential and $\Psi(x,t)=\Psi$ is the wavefunction in the coordinate representation, we perform a Madelung transformation.  

For the purpose of developing a quantum-hydrodynamic framework for investigating deviations from Born's rule, we employ the Madelung transformation, which reexpresses the wavefunction in terms of the probability density $\rho(\mathbf{r},t)$ and the phase $S(\mathbf{r},t)$. This polar transformation, rooted in the Bohmian formulation \cite{Norsen2017}, recasts the time-dependent Schr\"{o}dinger equation into hydrodynamic-like equations, including the continuity equation for $\rho$ and a quantum Hamilton–Jacobi equation for $S$.

	\begin{equation}
	\Psi=\sqrt{\rho}e^{iS/\hbar}.
	\end{equation}

\noindent
To advance our quantum-hydrodynamic framework for probing deviations from Born's rule, we introduce the total time derivative, or convective derivative, defined as

   \begin{equation}
       \frac{d}{dt}=\frac{\partial}{\partial t}+\mathbf{v}\cdot \nabla,
   \end{equation}
   
\noindent
where $\mathbf{v} = \frac{1}{m}\nabla S$ is the velocity field derived from the phase $S$ of the wavefunction via the polar transformation, and $m$ denotes the particle mass \cite{Norsen2017}. In the Bohmian formulation, this operator captures the rate of change of the probability density along quantum trajectories, thereby facilitating the derivation of the continuity equation,

\begin{equation}
     \dot{\rho}=-\rho \nabla \cdot\mathbf{v},
\end{equation}

\noindent
which admits a solution of the form

\begin{equation}
    \rho(t)=|\Psi(t)|^2=C e^{-\int ^{t} dt' \nabla\cdot \mathbf{v}}
\end{equation}

\noindent
where $C$ is a constant. The Hamilton–Jacobi equation reads as follows

 \begin{equation}
     \frac{\partial S}{\partial t}=-\frac{1}{2m}\left({\nabla S}\right)^{2}-V(\mathbf{r},t)-Q,
 \end{equation}

\noindent
from which we derive the equation of motion by applying the total time derivative,

 \begin{equation}
    m  \dot{\mathbf{v}}=-\nabla (Q+V)
    \end{equation}

 \noindent
    where 

    \begin{equation}
        Q=-\frac{\hbar ^{2}}{2m} \frac{\nabla ^{2 }\sqrt{\rho}}{\sqrt{\rho} }=-\frac{\hbar ^{2}}{4m}\left[\frac{\nabla ^{2}\rho}{\rho}-\frac{1}{2}(\nabla ln \rho)^{2}\right],
    \end{equation}

\noindent
the quantum potential which plays a pivotal role in our quantum-hydrodynamic framework, introducing non-local effects that distinguish quantum trajectories from classical dynamics. Unlike the classical potential $V$, the quantum potential $Q$ depends explicitly on the probability density.  

We now present the second-order deterministic differential equation for the probability density, a cornerstone of our conceptual framework for inducing deviations from Born's rule. This equation is derived by applying the total time derivative to the continuity equation introduced above and is written as follows.

\begin{equation}
    \ddot{\rho}=-\dot{\rho}\nabla \cdot \mathbf{v}-\rho \nabla \cdot \dot{\mathbf{v}}.
\end{equation}

\noindent
By construction, solutions of the continuity equation satisfy this second-order equation; however, the converse does not hold, as the latter permits non-unitary yet time-symmetric dynamics. These dynamics enable probability attenuation or amplification, leading to violations of Born's rule. We obtain a solution to the above deterministic equation by applying d’Alembert’s reduction method \cite{Billingham2008}. Utilizing a known solution corresponding to the quantum-equilibrium hypothesis, we find that

\begin{equation}
    \rho (t)=|\Psi(t)|^{2}\left(1+c'\int _{\tau}^{t}e^{-\int ^{t'}  dt''\nabla \cdot  \mathbf{v}} \frac{dt'}{|\Psi(t')|^{4}} \right).
\end{equation}

\noindent
We derived this form of solution, which includes a memory-kernel term that captures the temporal history of the wavefunction. It is possible to simplify the above expression by employing (5) and reducing it to

\begin{equation}
     \rho (t)=|\Psi(t)|^{2}\left(1+c\int _{\tau}^{t} \frac{dt'}{|\Psi(t')|^{2}} \right),
\end{equation}

\noindent
where

\begin{equation}
    c=\left(\frac{d\rho(t)}{dt}+|\Psi(t)|^{2}\nabla \cdot \mathbf {v}\right)\bigg|_{t=0},
\end{equation}

\noindent
a constant determined from initial conditions of $\rho$ and its first time derivative. The resulting probability density depends on the wavefunction’s values at previous times, reflecting its temporal structure and constitutes the main finding of this work. Unlike non-Markovian open quantum systems driven by an environment or thermal reservoir, our single-particle formulation relies solely on intrinsic nonlocal temporal effects. Near a wavefunction node, $\rho \rightarrow 0$ unless the divergence of the velocity field satisfies $\nabla \cdot \mathbf{v} = 0$, where it becomes ill-defined due to a singularity. Depending on the value of the constant $c$, $\rho$ may not be positive definite, in which case the situation may not be directly realizable or the conditions for its realization may not be fulfilled in principle \cite{Feynman1987}. Finally, an alternative formal expression for (11) is

  \begin{equation}
    \rho(t)=|\Psi|^{2}e^{c\int _{\tau}^{t} \frac{1}{\rho (t')}dt'}.
\end{equation}

\noindent
We may complement our exact solution by employing the first-order Dyson series in the Schr\"{o}dinger picture in order to derive an approximate expression for the probability density. For simplicity, we have assumed that the Hamiltonian at different times commutes, eliminating the need for chronological time-ordering in the perturbation series,

\begin{equation}
    \rho (x,t)  \approx \left [1+\frac{\left(\int _{\tau}^{t} \hat{H}dt'\right)^2}{\hbar ^2} \right]|\Psi _0|^2\left\{1+c\int \frac{dt'}{|\Psi _0|^2}\left[1-\frac{1}{\hbar^2}\frac{\left(\int _{\tau}^{t} \hat{H}dt'\right)^2|\Psi _0|^2}{|\Psi _0|^2}\right]\right\}.
\end{equation}

To investigate deviations from exponential decay at intermediate timescales, we connect our single-particle quantum-hydrodynamic framework to the Fock–Krylov formalism, utilizing its spectral approach to probe non-unitary dynamics. It is well known that quantum systems exhibit non-exponential decay at very short and long times due to the bounded energy spectrum \cite{Khalfin1958,Peres1980}. Here, we extend this analysis to intermediate timescales by applying a spectral Fourier transform to the wavefunction, which elucidates the impact of the energy (or frequency) distribution on the probability density,

\begin{equation}
    \rho (\omega)=|\Psi|^{2}(\omega)+c \int dt'' e^{i\omega t''} |\Psi(t'')|^{2}\int dt' \frac{ 1}{ |\Psi(t')|^{2}}.
\end{equation}

\noindent
To refine our analysis of non-exponential decay, we consider the spectral Fourier transform of the probability density in configuration space, which may exhibit poles and singularities due to the structure of the energy spectrum \cite{Khalfin1958}. Employing a standard prescription, we introduce an $i\varepsilon$ term in the denominator to regularize these singularities, ensuring a well-defined $\rho$. We then apply the Sokhotski-Plemelj theorem \cite{King2009} to evaluate the resulting integrals, yielding possible contributions from poles. Defining $g(\omega)$ such that

    \begin{equation}
    e^{i\omega t''} |\Psi(t'')|^{2}\int dt' \frac{ 1}{ |\Psi(t')|^{2}}=\frac{g(\omega)}{\omega- \omega _{0}\mp i\varepsilon},
\end{equation}

\noindent
and building on our spectral analysis, we derive an expression for the survival probability, which reflects the system’s persistence in its initial state, within the Fock-Krylov formalism to quantify non-exponential decay,

\begin{equation}
 p=\left|\int _{0}^{+\infty}e^{i\omega t}\left[|\Psi|^{2}(\omega)+c\left(\mathcal{P}\int _{-\infty}^{+\infty} d\omega g(\omega) \pm i\pi g(\omega _{0}) \right)\right] d\omega\right|^{2},
 \end{equation}

\noindent 
where $\mathcal{P}$ the principal part of the integral. It is evident that even when the Fourier transform of $|\Psi|^2$ has a Lorentzian profile, the spectral density may correspond to non-exponential decay.  

\textit{Modified interference patterns and examples of temporal nonlocality} Having established our single-particle quantum-hydrodynamic framework, we investigate modified interference patterns and temporal nonlocality through illustrative examples. First, we consider a particle in a box with a superposed wavefunction exhibiting quantum beats due to interference between energy eigenstates. The probability density, governed by our second-order differential equation with a memory-kernel term, deviates from the standard quantum expectation, reflecting nonunitary dynamics and non-exponential decay,  

\begin{equation}
    \Psi(x,t)=\frac{1}{\sqrt{L}}\left[ sin\left(\frac{\pi x}{L}\right)e^{-i\omega _{1}t}+sin\left(\frac{2\pi x}{L}\right)e^{-i\omega _{2}t}\right].
\end{equation}

\noindent
Substituting the above into (11), we find, with the help of integral tables \cite{Brychkov2010}, that  

  \begin{align}
      \rho =&\frac{1}{L}\left[sin^{2}\left(\frac{\pi x}{L}\right)+sin^{2}\left(\frac{\pi x}{L}\right)+2sin\left(\frac{\pi x}{L}\right)sin\left(\frac{\pi x}{L}\right)cos\delta\omega t \right] \nonumber  \\
  & \left[1+\frac{2c}{sin^{2}\left(\frac{\pi x}{L}\right)-sin^{2}\left(\frac{\pi x}{L}\right)}ln|\frac{tan\left(\frac{\delta \omega t' }{2}\right)A+B}{tan\left(\frac{\delta \omega t' }{2}\right)A-B}|\Bigg|_{t-\delta\tau}^{t}\right],
  \end{align}

\noindent
  where

\begin{subequations}
\begin{align}
  A=\left[ sin\left(\frac{\pi x}{L}\right)-sin\left(\frac{2\pi x}{L}\right)\right]^{2}, \\
   B=\left[ sin\left(\frac{\pi x}{L}\right)+sin\left(\frac{2\pi x}{L}\right)\right]^{2}.
\end{align}
\end{subequations}

\noindent
It is also possible to generalize the above for the case of two-particle interference, such as the Franson effect \cite{Franson1989} or experiments testing the Aharonov–Bohm effect with more than one particle \cite{Neder2007}.  

Next, we formulate a Clauser-Horne-Shimony-Holt (CHSH)-type inequality to test temporal nonlocal correlations, varying the final time and finite time delay. Unlike Leggett-Garg inequalities \cite{Emary2014}, which rely on consecutive measurements and dichotomic variables, our approach examines continuous temporal evolution. It can be readily demonstrated that the probability density given by (11) may violate this inequality.  

\begin{equation}
    \rho(t,\delta\tau)-\rho(t,\delta\tau')+ \rho(t',\delta\tau)+ \rho(t,\delta\tau')- \rho(t')- \rho(\delta\tau)\leq 0,
\end{equation}

\noindent
when one time interval exceeds another, $\delta \tau' > \delta \tau$, and the finite measurement duration is defined as the difference between the final time and a characteristic time delay. However, when the measurement duration approaches zero, $\delta \tau \rightarrow 0$, corresponding to an instantaneous measurement, such violations become impossible, as the temporal nonlocality diminishes.  

Our investigation into deviations from Born's rule reveals a novel framework for understanding quantum probability in non-equilibrium systems. By preserving the Schr\"{o}dinger equation and leveraging the Bohmian formulation, we demonstrate that finite-time nonlocal effects, driven by transit-interference phenomena and intrinsic memory effects, induce violations of the quantum-equilibrium hypothesis. Central to this approach is a deterministic second-order differential equation for the probability density, derived from the standard continuity equation, which captures nonlocal correlations. Unlike unitary quantum evolution, our model is time-symmetric but nonunitary, allowing for attenuation or amplification of the total probability. This nonunitarity may lead to violations of temporal nonlocal inequalities, offering a new perspective on finite-duration measurement problems and the time dwell problem in quantum tunneling.  

These findings challenge the universality of Born's rule, suggesting that its predictions may break down under specific temporal conditions. The non-exponential decay observed in unstable states at intermediate timescales underscores the role of memory effects and nonlocality in quantum dynamics. Our approach provides a theoretical foundation for probing these deviations, with potential applications in designing experiments to test temporal nonlocal inequalities in systems involving finite-duration measurements or tunneling phenomena. Future work could explore the implications of this framework for specific quantum systems, such as quantum optics or particle decay, and investigate experimental signatures of nonunitary evolution. By redefining the probabilistic structure of quantum mechanics, this study opens new avenues for understanding and manipulating non-equilibrium quantum phenomena.


\begin{thebibliography}{99}

\bibitem{Moshinsky1952}
M. Moshinsky, Diffraction in Time, \textit{Phys. Rev.} \textbf{88}, 625--631 (1952).
DOI: 10.1103/PhysRev.88.625.

\bibitem{Kleber1994}
M. Kleber, Exact solutions for time-dependent phenomena in quantum mechanics, \textit{Phys. Rep.} \textbf{236}, 331--393 (1994).

\bibitem{Norsen2017}
T. Norsen, \textit{Foundations of Quantum Mechanics: An Exploration of the Physical Meaning of Quantum Theory} (Springer, Cham, Switzerland, 2017).
ISBN: 9783319658667.

\bibitem{Billingham2008}
J. Billingham and A. C. King, \textit{Differential Equations: Linear, Nonlinear, Ordinary, Partial} (Cambridge University Press, Cambridge, UK, 2008).
ISBN: 9780521016872.



\bibitem{Feynman1987}
R. P. Feynman, Negative Probability, in \textit{Quantum Implications: Essays in Honour of David Bohm}, edited by B. J. Hiley and F. D. Peat (Routledge, London, 1987), pp. 235--248.
ISBN: 9780710208064.

\bibitem{King2009}
F. W. King, \textit{Hilbert Transforms: Volume 2} (Cambridge University Press, Cambridge, UK, 2009).
ISBN: 9780521517201.

\bibitem{Khalfin1958}
L. A. Khalfin, Contribution to the Decay Theory of a Quasi-Stationary State, \textit{Sov. Phys. JETP} \textbf{6}, 1053--1063 (1958) [Zh. Eksp. Teor. Fiz. \textbf{33}, 1371--1382 (1957)].


\bibitem{Peres1980}
A. Peres, Nonexponential Decay Law, \textit{Ann. Phys.} \textbf{129}, 33--46 (1980).
DOI: 10.1016/0003-4916(80)90330-9.


\bibitem{Franson1989}
J. D. Franson, Bell Inequality for Position and Time, \textit{Phys. Rev. Lett.} \textbf{62}, 2205--2208 (1989).
DOI: 10.1103/PhysRevLett.62.2205.

\bibitem{Neder2007}
I. Neder, Interference between two indistinguishable electrons from independent sources, \textit{Nature} \textbf{448}, 333--337 (2007).


\bibitem{Brychkov2010}
Y. A. Brychkov and O. Marichev, \textit{Handbook of Integrals and Series} (Chapman and Hall/CRC, Boca Raton, FL, 2010).
ISBN: 9781439828984.

\bibitem{Emary2014}
C. Emary, N. Lambert, and F. Nori, Leggett--Garg Inequalities, \textit{Rep. Prog. Phys.} \textbf{77}, 016001 (2014).
DOI: 10.1088/0034-4885/77/1/016001.




\end{thebibliography}
\end{document}